# Non-Contact linear slider for cryogenic environment

José-Luis Pérez-Díaz*, Juan Carlos García-Prada*, Efrén Díez-Jiménez*, Ignacio Valiente-Blanco*, Berit Sander*, Lauri Timm*, Juan Sánchez-García-Casarrubios*, Javier Serrano**, Fernando Romera**, Heribert Argelaguet-Vilaseca** and David González-de-María**

*Dto. de Ingeniería Mecánica, Universidad Carlos III de Madrid, Butarque, 15, E-28911 Leganés, Spain
** LIDAX, Cristobal Colon 16. E-28850 Torrejón de Ardoz, Spain.

**Abstract**

A non-contact linear slider based on stable superconducting magnetic levitation with a long permanent magnet as a slider and two fixed superconducting disks which define the slide way has been designed, built and tested. The slider can be moved stably along a stroke of ±11.5 mm by supplying a low current in a coil located at the end of the stroke while the levitation remains stable providing a reliable mechanism for linear displacement in a cryogenic environment. The response is linear with a sensitivity of 52±2 µm/mA for displacements lower than 6 mm. Run out, pitch, yaw and roll have been measured demonstrating an overall good performance. In all cases the measured hysteresis was lower than 250 µm and the measured run out was also lower than 250 µm both for Y and Z axis.

Roll and yaw were always below 300 µrad, that is one order of magnitude lower than the pitch (4500 µrad).

* Corresponding author. Tel.: +34 916249912; fax: +34 916249912.
E-mail address: jlperez@ing.uc3m.es

# 1. Introduction

There is an increasing demand for very precise mechanisms able to work in cryogenic environments for use in the aerospace industry both in space or on earth, optical communication, and bio-medical precision industries[1,2]. Ultraprecision instruments, equipped with the most sensitive devices and sensors require very low temperature, sometimes close to 4K. These instruments, like interferometers, also need mechanisms for very accurate positioning.

In this context, conventional mechanisms are usually based in gears and present severe tribological problems like backlash and cold spots, fatigue or wearing. Only solid lubrication is available at very low temperature with a reasonable good performance [3,4,5].

Superconducting magnetic levitation provides a new tool for mechanical engineers to design non-contact mechanisms solving all the tribological problems associated with contact at very low temperature[6,7].

Forces between a permanent magnet and a superconductor can be both repulsive and attractive, providing stable interaction without direct contact. This is particularly true for type II superconductors being at the mixed state.

In the last years, some modeling tools have been developed in order to help engineers to design mechanism based on the superconducting magnetic levitation [8,9]. Not only forces but also torques can be calculated in the Meissner state in good agreement with experimental data [10,11,12,13]. Methods of calculation in the mixed state have been reported as well and some of them had been experimentally validated [14,15].

Although there has been a great interest in magnet superconducting bearings [16,17], it has been paid very little attention on superconducting magnetic levitation sliders. Only a kind of short-range micro-conveyor based in superconducting magnetic levitation has been proposed [18]. As far as we know, nobody has reported any linear slider based on superconducting magnetic levitation able to displace a mobile part with high accuracy within a stroke longer than few millimeters.

We have designed, built and tested a prototype of a non-contact linear slider based on superconducting magnetic levitation using type II superconductors and a permanent magnet with an easy position control system able to move accurately the slider within a long stroke.

# 2. Description of the mechanism

The mechanism studied in this paper is mainly composed of a stator with two superconducting polycrystalline $YBa_2Cu_3O_{7-x}$ disks (1) and a slider (2) with a long permanent magnet as shown in Fig.1. The superconductor disks (1) and the permanent magnet (2) form a kinematic pair such that the slider levitate stably over the superconductors and can freely move along the X axis (see Fig.1) without any contact.

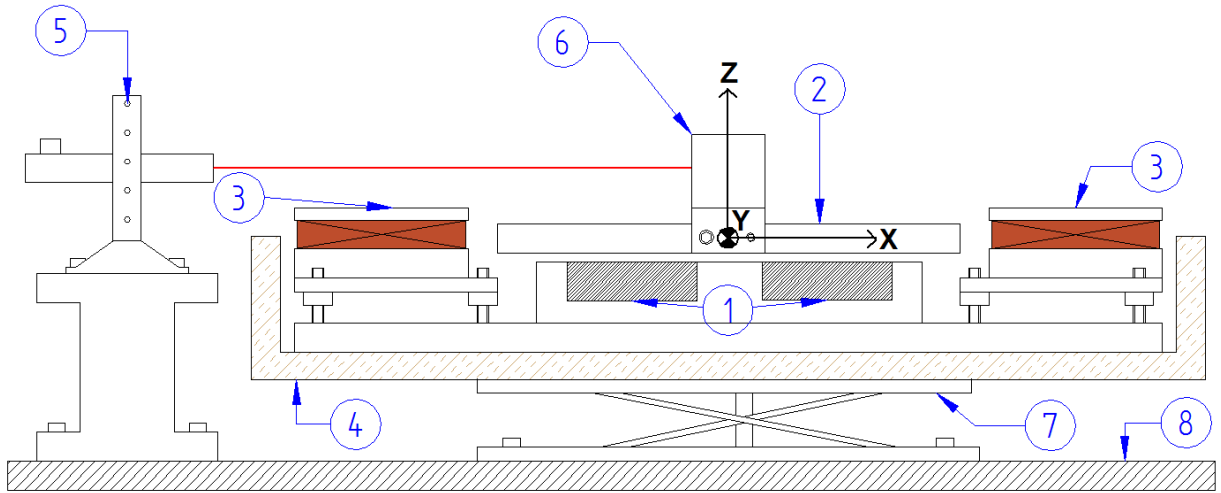

Fig. 1. Sketch of the experimental set-up: 1 YBaCuO superconductor disks; 2 Slider permanent magnet; 3 Coils, 4 liquid nitrogen vessel; 5 Collimator; 6 Polished aluminum mirror cube; 7 lab-jack stand and 8 optic table.

The height of levitation is established before cooling down and it will be called field cooling height (FCH). FCH is the distance between the top surface of the superconductors and the bottom surface of the permanent magnet. For all the experiments reported in this work FCH is equal to 3 mm. Once the superconductor disks are cooled down below their critical temperature ($Tc$), the sliding kinematic pair is established providing stable levitation to the slider as long as the superconductor disks remain at a low enough magnetic field.

In this case the superconductor disks are cooled down using liquid nitrogen (77 K) below their critical temperature ($Tc$=90 K). While being cooling down, they are immersed in the magnetic field generated by the permanent magnet as shown in Fig.2.

The flux trapping effect makes the permanent magnet to levitate stably over the superconductor [16]. Then, if the position of the permanent magnet changes, the magnetic field seen by the superconductor changes and drag forces appear trying to restore the initial position and therefore the initial magnetic field [19].

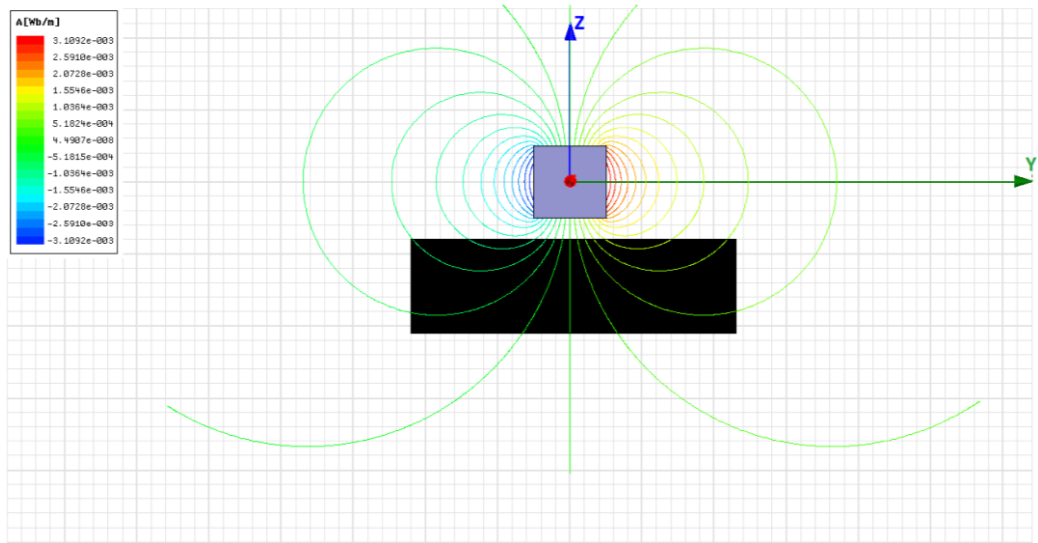

Fig. 2. Magnetic Flux Lines represented in a cross section.

However, due to the design of this mechanism, the magnetic field seen by the superconductor disks remains invariant with respect to the X position of the slider. Therefore, the permanent magnet can slide with an extremely low resistance to its movement.

Nevertheless, when any of the ends of the permanent magnet is close enough to any of the edges of the superconductor disks, the symmetry is broken and the end of the stroke is reached. In fact the finiteness of the length of the slider breaks the translational symmetry of the kinematic pair creating a stable position. In this paper we demonstrate that the effect of the edges breaking symmetry is a second order effect that can be used to move stably and simply the slider. The forces needed to slide are much lower than those required to get it off the "guide".

We can therefore say that guidance or "slide way" is provided by the two superconductor disks with a stroke determined by the distance between the end of the permanent magnet and the superconductor disks edges. See Fig.3.

In order to change the X position of the slider, a pair of coils (3) (see Fig.1) placed at the end of the stroke were used. By varying the electrical current circulating in these coils, X position of the slider can be modified precisely.

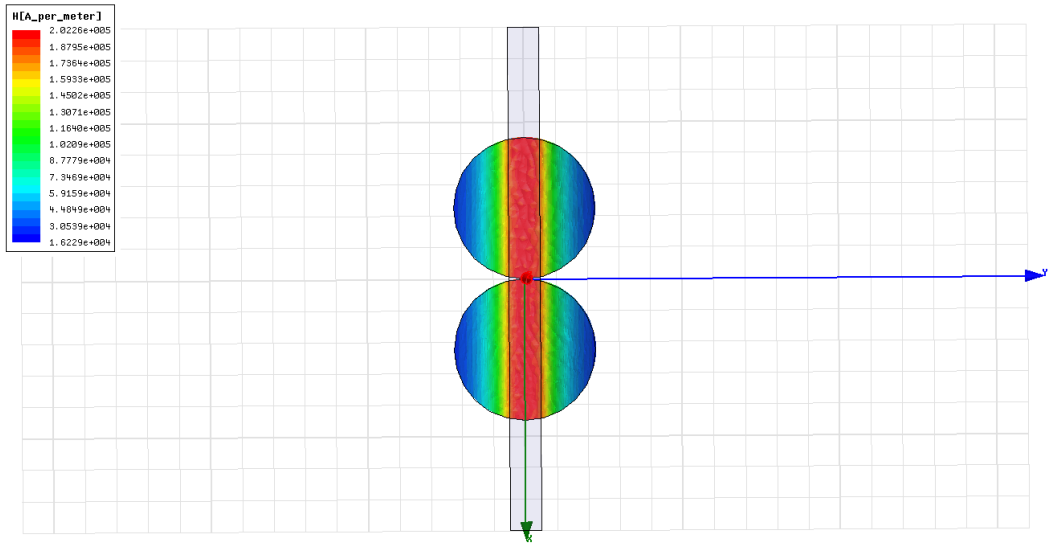

Fig. 3. Magnetic Field B[T] magnitude on superconductor surfaces. The "slide way" can be clearly seen.

The translational symmetry of the magnetic field in the mechanism presented in this work is, therefore, similar to the rotational symmetry that has been used in several magnetic levitation devices as superconducting magnetic bearings [20,21,22] where usually the rotor is able to spin but not to move in other directions because a drag force appears [7].

## 3. Experimental set-up and procedure

A prototype was built to test the precision of guidance and stability of positioning of this mechanism. Two polycrystalline $YBa_2Cu_3O_{7-x}$ disks with a 45 mm nominal diameter and 13 mm height were used as superconductor bulks in the stator (see Fig.1). These disks were located in a plate made of diamagnetic aluminum conveniently shaped to keep them in position.

The slider was a 160 mm long permanent magnet (2) made of NdFeB with a coercivity of 900 kA/m and a remanence of 1.30 T. The cross section is a 10 mm side square. It was placed symmetrically over the superconductors with its symmetry axis carefully aligned with the X axis and the magnetization direction parallel to the Z axis. An image of the device is shown in Fig.4.

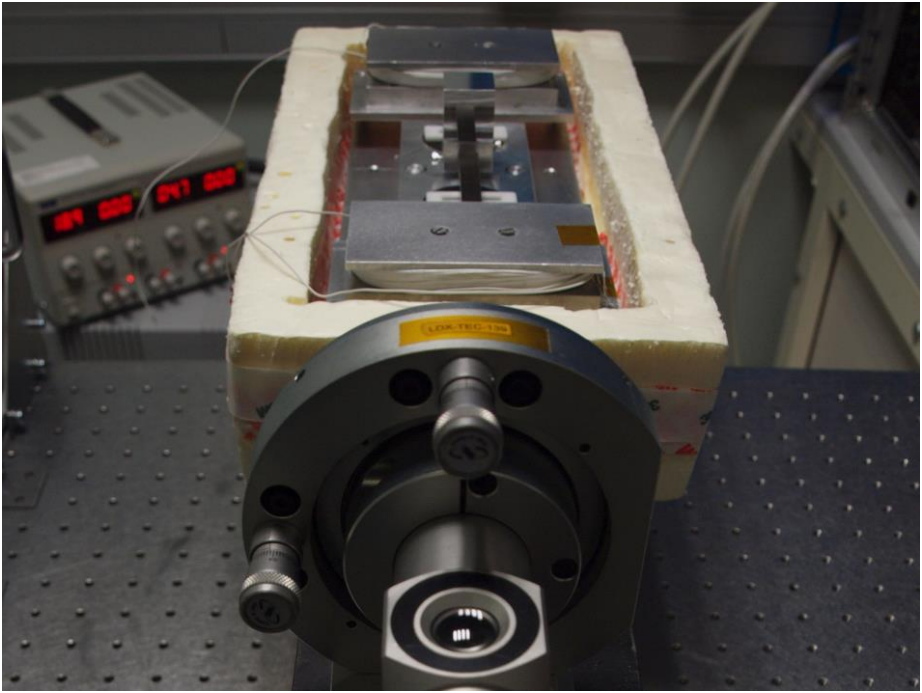

Fig.4. View of the prototype of the non-contact slide way in the liquid nitrogen container.

The prototype was tested in a bath of liquid nitrogen at ambient pressure to guarantee the temperature of the superconducting disks below its critical temperature (*Tc*). The permanent magnet was not in direct contact with the liquid nitrogen at any time. An extruded polystyrene vessel (4) was used to contain both the liquid nitrogen and the prototype (see Fig.1).

Scales were used to measure the displacements along the three axes with a precision of 250 μm. A collimator (5), with an accuracy of 0.1μrad, was used to measure the three rotations of the permanent magnet (pitch, yaw and roll). An optic mirror cube (6) made of polished aluminum was fixed to the slider (2) to reflect the laser beam from the collimator. A lab-jack stand (7) was also used to adjust the relative height of the whole prototype with respect to the laser beam from the collimator. Finally, the prototype and other auxiliary components were installed on an optic table (8) which provides good horizontality and vibration insulation.

A power supply with two independent channels was connected to the coils (3). In addition, two respective digital current meters with 10μA resolution were used to measure the current in the coils. This system is therefore able to supply a different current to each of the two coils to change the X position of the slider.

All the measurements –except for the X rotation- were performed as shown in Fig.1. The measurement of the rotation angle around the X axis needed a second configuration with the collimator parallel to the Y axis. This second configuration allows measuring the X rotation while the laser beam points on the optic mirror cube. As this cube is only one inch wide this is, more or less, the range of X position for which X rotation may be measured.

## 4. Results and discussion

The prototype was cooled down to its working temperature of 77 K and the slider remained at a stable position. Then the X position was smoothly varied by supplying a current to the coils: The movement was demonstrated to be strongly stable and the control of the position could be carried out just by direct control of the current supplied to the coils. For all the tests performed the permanent magnet or slider is moved from the central position (X=0) along the X axis and back to the central position. This is an efficient way to measure possible hysteretic behavior.

The system of coordinates used has the X axis parallel to the direction of sliding and the Z axis vertically oriented with the origin of coordinates located at the center of mass of the slider. The Y axis is horizontally directed as shown in Fig.1.

In all cases the measured hysteresis was lower than 250 µm and the measured run out was also lower than 250 µm both for Y and Z axis.

*4.1 X position vs. current in the coil*

The X position of the slider along its path vs. the current supplied into the coil is shown in Fig.5. Data were obtained both increasing and decreasing the current throuhg the coils.

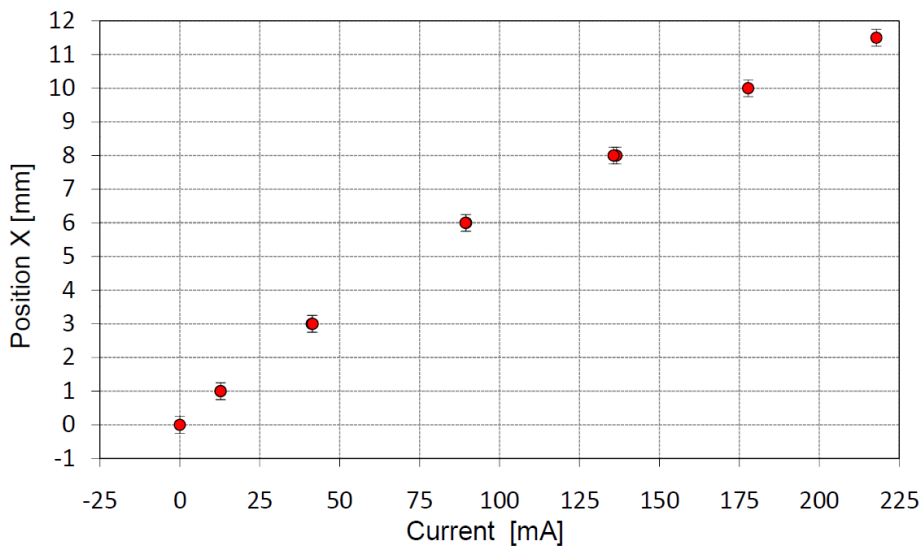

Fig. 5.  X position of the slider vs. current in the coil.

The displacement of the slider presents a pretty linear dependence on the current from 0 to 6 mm.  Only when one of the ends of the permanent magnet approaches to any edge of the superconductor disks, the symmetry is increasingly broken and the resistance to the movement of the slider also increases. Therefore, the current needs to be increased more than linearly.  It has to be noted that relatively large displacements are reached with very low current (around 52±2 µm/mA) for the linear stage.

*4.2 Pitch*

Pitch (or rotation around the Y axis) of the slider vs. X position along the "slide way" is shown in Fig. 6. The standard deviation is between 10 and 20 µrad.

Good linearity can be observerd for the full stroke. This can be related to effects of gravity on the mechanism. As the center of mass of the slider moves from the central position the lift forces become unbalanced. Nevertheless, the pitch remains limited due to the stiffnes of the levitating forces involved. A hysteresis of the order of 700 µrad appears in the origin.

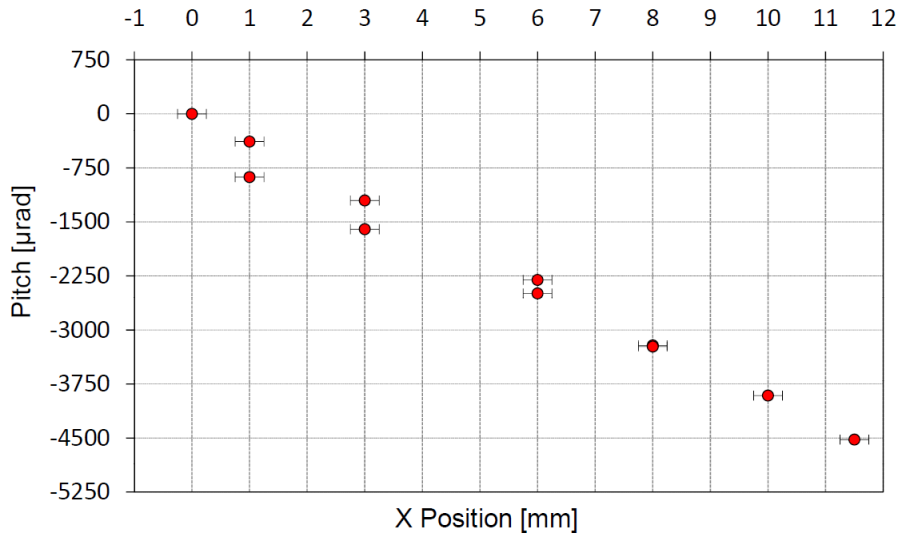

Fig. 6. Pitch vs. X position of the slider.

*4.3 Yaw*

Yaw (or rotation around the Z axis) of the slider vs. X position along the "slide way" is shown in Fig. 7. The standard deviation is between 10 and 20 µrad. Yaw is supposed to appear as a result of a missalignment between the line from the center of mass to the axis of the coil and the axis of the permanent magnet. A hysteresis of the order of 20 µrad is observed in the origin. The yaw is always one order of magnitude lower than the pitch.

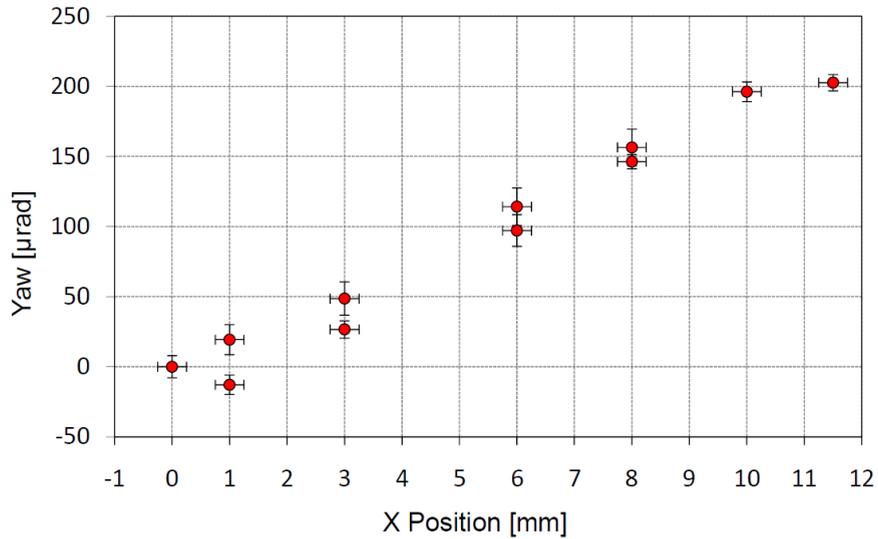

Fig. 7. Yaw vs. X position of the slider.

*4.4 Roll*

Roll (or rotation around the X axis) of the slider vs. X position along the "slide way" is shown in Fig. 8. The standard deviation is between 10 and 20 μrad. To characterize the roll the collimator was fixed in a lateral position with the laser beam along the Y axis. The measurable range of displacement is then limited by the size of the mirror on the slider (25 mm). Roll is supposed to be caused by missalignements between the axis of the coils and the magnetization axis of the permanent magnet in the slider. Similarly to the yaw, the roll is always one order of magnitude lower than the pitch.

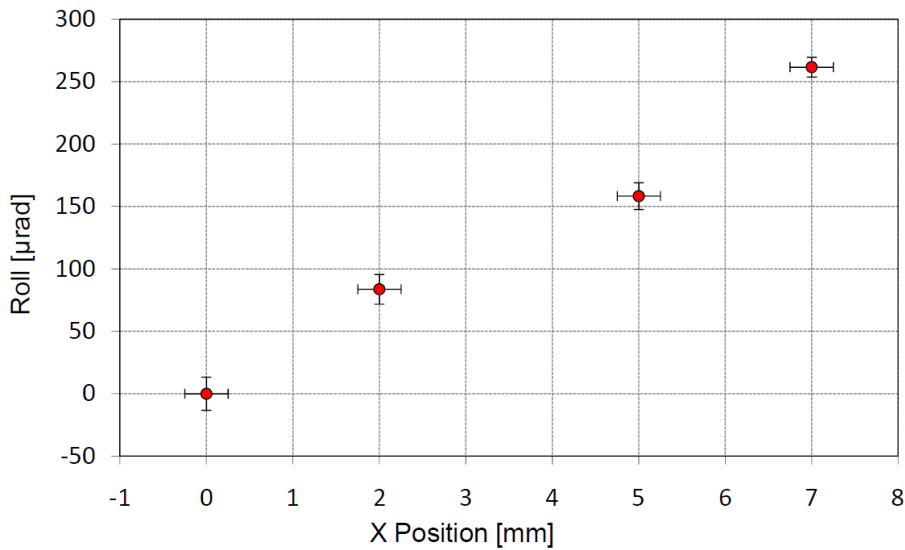

Fig. 8. . Roll vs. X position of the slider.

## 5. Conclusions

We have designed, built and tested a non-contact linear slider based on stable superconducting magnetic levitation, with a long permanent magnet as a slider and two fixed superconducting disks which define the slide way. As the magnetic field seen by the superconducting disks remains almost invariant with respect to the X position of the slider, the permanent magnet can slide with an extremely low resistance.

We have checked that the forces needed to slide are much lower than those required to get it off the slide way. The slider can be moved stably supplying a low current in a coil located at the end of the stroke while the levitation remains stable. The maximum stroke was of 11.5 mm. There is a linear dependence with a slope of 52±2 µm/mA for displacements lower than 6 mm.

In all cases the measured hysteresis was lower than 250 µm and the measured run out was also lower than 250 µm both for Y and Z axis.

Roll and yaw were always below 300 µrad, that is one order of magnitude lower than the pitch (4500 µrad). This is due to the fact that gravity directly affects the pitch, but roll and yaw are affected only by initial missalignements.

Finally, we have proved out that a non-contact slider based on superconducting levitation may be controlled simply and precisely by adjusting the current in a coil. This, therefore, provides a reliable mechanism for linear positioning able to work in cryogenic conditions with a good performance.

**Acknowledgment**

This work has been partially funded by Dirección General de Economía, Estadística e Innovación Tecnológica, Consejería de Economía y Hacienda, Comunidad de Madrid, ref. 12/09.

**References**

[1] ESA, European non-dependence on critical space technologies: EC-ESA-EDA list of urgent actions for 2009. 2009.
[2] S. Devasia, E. Eleftheriou, S.O.R. Moheimani, A survey of control issues in nanopositioning, Control Systems Technology, IEEE Transactions On. 15 (2007), 802–823.
[3] Y.L. Ostrovskaya, T. Yukhno, G. Gamulya, Y.V. Vvedenskij, V. Kuleba, Low temperature tribology at the B. Verkin Institute for Low Temperature Physics & Engineering (historical review), Tribology International. 34 (2001), 265–276.
[4] A. Trautmann, C.R. Siviour, S.M. Walley, J.E. Field, Lubrication of polycarbonate at cryogenic temperatures in the split Hopkinson pressure bar, International Journal Of Impact Engineering. 31 (2005), 523-544.
[5] N. Fleischer, M. Genut, L. Rapoport, R. Tenne, New nanotechnology solid lubricants for superior dry lubrication, Proceedings Of the 10th European Space Mechanisms and Tribology Symposium, 2003, 65 - 66.
[6] J.L. Pérez Díaz and E. Díaz Jiménez, Foundations of Meissner superconductor magnet mechanisms engineering Superconductivity - Theory and Applications, 2011.
[7] F.C. Moon, PZ. Chan, Superconducting Levitation: Applications to bearings and magnetic transportation, Wiley-VCH, Germany, 1994.
[8] J.L. Perez-Diaz, J.C. Garcia-Prada, Finite-size-induced stability of a permanent magnet levitating over a superconductor in the Meissner state, Applied Physics Letters. 91 (2007) 142503.
[9] E. Diez-Jimenez, J.-L. Perez-Diaz, J.C. Garcia-Prada, Local model for magnet–superconductor mechanical interaction: Experimental verification, Journal Of Applied Physics. 109 (2011) 063901.
[10] J.L. Pérez Díaz and J.C. García-Prada, Universal model for Meissner levitation, International Review Of Mechanical Engineering. (2008) 6-11.
[11] J.L. Perez-Diaz, J.C. Garcia-Prada, Interpretation of the method of images in estimating superconducting levitation, Physica C: Superconductivity. 467 (2007) 141-144.
[12] Ignacio Valiente-Blanco, Efren Diez-Jimenez, J.-L.P.-D., Alignment effect between a magnet over a superconductor cylinder in the Meissner state, Journal Of Applied Physics. 109 (2011).
[13] E. Diez-Jimenez, J.-L. Perez-Diaz, Flip effect in the orientation of a magnet levitating over a superconducting torus in the Meissner state, Physica C: Superconductivity. 471 (2011) 8-11.

[14] A.A. Kordyuk, Magnetic levitation for hard superconductors, Journal Of applied Physics. 83 (1998) 610–612.

[15] D.H.N. Dias, E.S. Motta, G.G. Sotelo, R. de Andrade, R.M. Stephan, L. Kuehn, et al., Simulations and Tests of Superconducting Linear Bearings for a MAGLEV Prototype, IEEE Transactions On Applied Superconductivity. 19 (2009) 2120-2123.

[16] J.R. Hull, Superconducting bearings, Superconductor Science and Technology, 13 (2000) 15.

[17] S.O. Siems, W.R. Canders, Advances in the design of superconducting magnetic bearings for static and dynamic applications, Superconductor Science and Technology. 18 (2005) S86-S89.

[18] F. Tetsuhiko, I. Hiroyuki, A micro conveyance actuator based on superconducting magnetic levitation, Transactions Of the Institute Of Electrical Engineers Of Japan. 119- E (1999) 417-23.

[19] AB. Riise, TH. Johansen, The vertical magnetic force and stiffness between a cylindrical magnet and a high-Tc superconductor, Physica C: Superconductivity. 234 (1994) 108-114.

[20] C.K. McMichael, K.B. Ma, M. a Lamb, M.W. Lin, L. Chow, R.L. Meng, et al., Practical adaptation in bulk superconducting magnetic bearing applications, Applied Physics Letters. 60 (1992) 1893.

[21] T.M. Mulcahy, J.R. Hull, K.L. Uherka, R.G. Abboud, J.H. Wise, D.W. Carnegie, et al., A permanent-magnet rotor for a high-temperature superconducting bearing, IEEE Transactions On Magnetics. 32 (1996) 2609-2612.

[22] Z. Xia, Q.Y. Chen, K.B. Ma, C.K. McMichael, M. Lamb, R.S. Cooley, et al., Design of superconducting magnetic bearings with high levitating force for flywheel energy storage systems, IEEE Transactions On Appiled Superconductivity. 5 (1995) 622-625.